\documentclass[11pt]{article}

\usepackage{latexsym,amsmath,epsfig}

\DeclareFontFamily{OT1}{rsfs10}{}
\DeclareFontShape{OT1}{rsfs10}{m}{n}{ <-> rsfs10 }{}
\DeclareMathAlphabet{\mathscript}{OT1}{rsfs10}{m}{n}

\numberwithin{equation}{section}

\newcommand{\be}{\begin{equation}}
\newcommand{\ee}{\end{equation}}

\newcommand{\bea}{\begin{eqnarray}}
\newcommand{\eea}{\end{eqnarray}}
\newcommand{\ba}{\begin{array}}
\newcommand{\ea}{\end{array}}

\newcommand{\ns}{\normalsize}

\begin{document}

\begin{titlepage}

\title{
\hfill{\ns hep-th/0310103\\[2cm]}
{\LARGE Kink-boundary collisions in a two dimensional scalar field theory}\\[1cm]}
\setcounter{footnote}{0}
\author{{\ns\large
 Nuno D. Antunes\footnote{email: n.d.antunes@sussex.ac.uk}~,
 Edmund J.~Copeland\footnote{email: e.j.copeland@sussex.ac.uk}~,
\setcounter{footnote}{3}
 Mark Hindmarsh\footnote{email: m.b.hindmarsh@sussex.ac.uk}~~and
 Andr\'e Lukas\footnote{email: a.lukas@sussex.ac.uk}} \\[0.8em]
      {\ns Centre for Theoretical Physics,
     University of Sussex}\\[-0.2em]
      {\ns Falmer, Brighton BN1 9QJ, United Kingdom}}


\maketitle

\vspace{1cm}

\begin{abstract}

In a two-dimensional toy model, motivated from five-dimensional heterotic
M-theory, we study the collision of scalar field kinks with boundaries.
By numerical simulation of the full two-dimensional theory, we find that
the kink is always inelastically reflected with a model-independent
fraction of its kinetic energy converted into radiation. We show
that the reflection can be analytically understood as a fluctuation
around the scalar field vacuum. This picture suggests the possibility
of spontaneous emission of kinks from the boundary due to small
perturbations in the bulk. We verify this picture numerically by
showing that the radiation emitted from the collision of an initial single kink
eventually leads to a bulk populated by many kinks. Consequently,
processes changing the boundary charges are practically unavoidable
in this system. We speculate that the system has a universal final
state consisting of a stack of kinks, their number being determined
by the initial energy.
\end{abstract}

\thispagestyle{empty}

\end{titlepage}


\section{Introduction}
\label{introduction}
In Ref.~\cite{Antunes:2002hn} we presented a five-dimensional
brane-world model, closely related to five-dimensional heterotic
M-theory~\cite{Lukas:1998yy,Ellis:1998dh,Lukas:1998tt,Lukas:1998hk,Brandle:2001ts}, where
M-theory five-branes are modelled as kink-solutions of a bulk
scalar field theory. We were particularly interested in studying
collisions of such kinks with the boundaries, a process which,
in the original M-theory model, may lead to a topology-changing
so-called small-instanton transition~\cite{Witten:1996gx,Ganor:1996mu}
due to absorption of the five-brane.
The analysis of Ref.~\cite{Antunes:2002hn} was mainly performed
in a moduli space approximation. We concluded that the colliding
kink was absorbed by the boundary, however, its final fate could
not be determined due to a break-down of the moduli space approach.

In this paper, we will focus on an even simpler, two-dimensional model
which captures the essential features of its five-dimensional cousin. 
The model consists of a ($1+1$)-dimensional scalar field theory for
a single real scalar and an associated potential $V$ with (at least)
two minima to allow for kink solutions. The spatial direction is taken
to be a line segment with the boundary conditions being provided by the
``superpotential"~\cite{DeWolfe:1999cp} $W$ associated to $V$. 
To simplify the problem we have not included gravity and any
other gravity-like fields which were present in the five-dimensional
model~\cite{Antunes:2002hn}.

Our main goal is to determine the final outcome of the kink-boundary
collision using this simple two-dimensional model. In particular,
we would like to clarify whether the kink is indeed absorbed or,
rather, reflected by the boundary. While the former process would
lead to a change in the boundary charge (as measured by the superpotential
value) and the number of kinks present in the bulk, the latter process
would conserve those numbers. We will mainly approach this problem as
one posed within  relatively simple two-dimensional scalar field
theories whose basic properties we are investigating, although the
possible relation to M-theory and topology-changing phenomena is in
the back of our minds.     

The plan of the paper is as follows. In the next Section, we will
set up the model and derive its basic kink solution. Section 3
reviews the moduli space description of kink evolution and kink-boundary
collision. The perturbation spectrum on the kink background is analysed
in Section 4. In Section 5, we present a full numerical analysis of
the two-dimensional model to determine the final outcome of the collision
process. An analytic interpretation of the numerical results is given
in Section 6. In Section 7, we present a modified moduli space picture
which incorporates some, although not all, features of the collision
process. We conclude in Section 8 by summarising and presenting some
results about the long-time evolution of the system.

\section{The model}
\label{model}

In this section, we will set up the $1+1$-dimensional model which will
be the basis of this paper. This model is designed to capture the
essential features of five-dimensional heterotic
M-theory~\cite{Lukas:1998yy} and the related defect model presented in
Ref.~\cite{Antunes:2002hn} and, at the same time, provide the simplest
setting for studying the collision of a kink with a boundary.

Time and spatial coordinate are denoted by $t$ and $x$, respectively,
where the latter is taken on a circle, that is, in the range $x\in
[-L,L]$ (with the endpoints of the interval identified) subject to
the $Z_2$ orbifolding generated by $x\rightarrow -x$. The resulting
one-dimensional orbifold $S^1/Z_2$ has two fixed points at $x=x_1=0$
and $x=x_2=\pm L$ and points $x$ and $-x$ are identified. Instead of
working with the full orbicircle, we, therefore, can (and frequently
will) use the ``downstairs'' picture where $x$ is restricted to the
interval $x\in [0,L]$. The fixed points at $x_i$, where $i=1,2$, can
now be thought of as spatial boundaries.

The field content of our model consists of a single real scalar field $\phi$
with a potential $V=V(\phi )$. Gravity and gravity-like fields will
not be considered here since they are not essential to analyse the
kink-boundary collision. We require the potential to have at least
two different minima (with vanishing potential values) in order for
kink solutions to exist. We also define the ``superpotential'' $W$ by
\begin{equation}
 W'=\sqrt{2V(\phi )}\; .
\end{equation}
where here and in the following the prime denotes the derivative with
respect to $\phi$. We will be interested in actions of the type
\begin{equation}
 S =\int {\rm dt} \int_{-L}^L {\rm dx}\,\left\{ \frac{1}{2}
     \left(\frac{\partial\phi}{\partial t}\right)^2 - 
     \frac{1}{2}\left(\frac{\partial \phi}{\partial
     x}\right)^2 - V(\phi) - 2\delta(x-0)\, W(\phi)+ 2\delta(x-L)\,
     W(\phi) \right\}
  \label{Action}
\end{equation}
The equations of motion and boundary conditions in the boundary picture
are then given by
  \begin{equation}
  \frac{\partial^2 \phi}{\partial t^2}=\frac{\partial^2 \phi}{\partial
   x^2}-V'(\phi),\qquad\qquad
   \left.\frac{\partial \phi}{\partial x}\right|_{x_i}=\left. 
   W'(\phi)\right|_{x_i}\; .
 \label{EOM}
\end{equation}
For the numerical simulations and for concreteness we will often use
the quartic potential
\begin{equation}
  V(\phi)=\frac{1}{8}g^2 (\phi^2-v^2)^2\label{quartic}
\end{equation}
with associated superpotential
\begin{equation}
 W = - \frac{1}{2} g\left(\frac{1}{3}\phi^3 - \phi v^2\right)\; .
 \label{quarticW}
\end{equation}
The form of the boundary terms in the action~\eqref{Action}
has been chosen to facilitate the existence of BPS solutions.
This can be understood by focusing on static configurations
$\phi =\phi (x)$. For such configurations, the action can be
written as
\begin{equation}
 S\sim\int_0^Ldx\left(\frac{d\phi}{dx}-W'\right)^2\; .
\end{equation}
Note that the boundaries terms which arise from partial integration
when converting the action~\eqref{Action} into the above form are
precisely cancelled by the original boundary terms in~\eqref{Action}.
Hence, we find static solutions $\phi =\phi (x)$ of our action
satisfy the first order equation
\begin{equation}
 \frac{d\phi}{dx}=W'\; .\label{EOM1}
\end{equation}
Conversely, solutions to this first order equation clearly satisfy the
boundary conditions as well as the second-order equation of motion in
\eqref{EOM}, the latter by virtue of the relation between $V$ and $W$.

{}From eq.~\eqref{EOM1}, the solution $\phi_{\rm K}$ of a kink can in
general be written as
\begin{equation}
 \phi_K(x) = f(x-Z)\; , \qquad f^{-1}(\phi )=\int_{\phi_0}^\phi
             \frac{d\tilde{\phi}}{W'(\tilde{\phi})}\; .\label{gensol}
\end{equation}
Here we take $\phi_0$ to be the $\phi$ value at the core of the kink,
corresponding to the maximum of the potential. Then, the constant $Z$ can
be interpreted as the position of the kink's core as long as
$Z\in [0,L]$. For $Z$ outside this range, the core of the kink is
no longer within the physical part of space and $Z$ merely indicates
the ``virtual'' position of the core. 

For the quartic potential~\eqref{quartic} the kink solution is given by
\begin{equation}
 \phi_{\rm K}(x)=v\; \tanh\left(\frac{x-Z}{2L_{\rm K}}\right)
 \label{profile}
\end{equation}
where the width $L_{\rm K}$ of the kink is given by
\begin{equation}
 L_{\rm K} = \frac{1}{gv}\; .
\end{equation} 

As long as kinks are sufficiently away from the boundaries, the
field $\phi$ will be close to minima of its potential $V$ on the
boundaries. Then, the boundary superpotential takes values from a
discrete set consisting of the values $W$ assumes at the minima of
the potential $V$. These discrete values can be interpreted as
boundary charges. They are analogous to the topological boundary
charges which appear in the related M-theory
model~\cite{Lukas:1998yy,Antunes:2002hn}. In order 
to consistently maintain this interpretation, the superpotential must 
be monotonically increasing, and so Eq.~(\ref{quarticW}) is only valid 
for $-v < \phi <v$.

We note that our boundary conditions are related  to the family considered in 
\cite{IntegBound}. These papers studied integrable ($1+1$)-dimensional 
field theories on the half line, including the sine-Gordon theory, 
with a two-parameter set of boundary conditions 
$\phi' = MW'(\phi - \phi_0)$ at the origin.  The fact that these boundary conditions 
do not spoil the integrability generated considerable interest.

\section{Moduli space evolution}
\label{moduli}

In the previous section, we have presented a one-parameter family of
static BPS kink solutions for our theory~\eqref{Action} which is
labelled by the parameter $Z$. Solutions exist for all values $Z\in
[-\infty ,\infty ]$ but only for $Z\in [0,L]$ does $Z$ correspond to
the position of the kink's core. Otherwise, the kink's core is located
``outside'' and only its tail remains within the physical part of
space. Values $Z=0$ or $Z=L$ correspond to a kink with its core being
exactly located at one of the boundaries.

We would now like to consider physical motion in this moduli space of
kinks by promoting $Z$ to a slowly-varying function of time,
$Z=Z(t)$. A time evolution $Z\rightarrow 0_+$ ($Z\rightarrow L_-$)
then describes the collision of the kink with the boundary at
$x=x_1=0$ ($x=x_2=L$) which is the process we would like to study in
this paper. For definiteness, we will focus on collisions with
the boundary at $x=0$ in the following but the conclusions, of course,
apply to collisions with the other boundary at $x=L$ as well. 

The one-dimensional effective theory $S_{\rm eff}$ for the modulus
$Z=Z(t)$ can be obtained by inserting the kink-solution~\eqref{gensol}
into the action~\eqref{Action} and integrating over the spatial
coordinate. This results in
\begin{equation}
  S_{\rm eff}=\int\ dt\, F(Z)\, \dot{Z}^2,
  \label{Seff}
\end{equation}
where the function $F=F(Z)$ is given by
\begin{equation}
 F(Z) = W(f(L-Z))-W(f(-Z))\; .
\end{equation}
The last equation means that $F$ is the difference of the boundary
superpotentials evaluated on the kink solution. From this
interpretation the qualitative behaviour of $F$ can be easily
inferred. For $Z\in [0,L]$ and well away from the boundaries
(that is, away by more than the width of the kink) the field
$\phi_{\rm K}$ is very close to neighbouring minima of the potential on
the two boundaries. More precisely, $\phi_{\rm K} (x=0)$ is very close
to the minimum with the smaller superpotential value and $\phi_{\rm K}
(x=L)$ is very close to the neighbouring minimum with the larger
superpotential value. This means, for such values of $Z$ the function
$F(Z)$ is finite positive and approximately constant. On the other hand, for
$Z<0$ or $Z>L$ and away from the boundary, $\phi$ is very close to the
same minimum for both boundaries which means that $F(Z)$ is
approximately zero (although still positive). 

\begin{figure}
\centering
\includegraphics[height=9cm,width=9cm, angle=0]{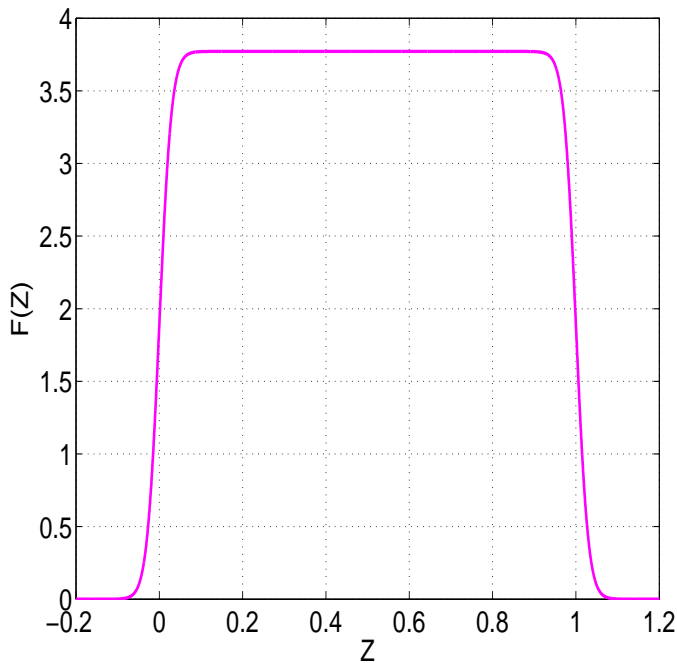}
\caption{\emph{The function F(Z) for an interval with length $L=1$ 
               and a kink width of $L_{\rm K}=\frac{1}{32\sqrt{2}}$.}}
\label{fig1}
\end{figure}

The function $F$ can be computed for any given potential $V$ and associated
superpotential $W$ to confirm this expectation. For the quartic
potential~\eqref{quartic}, the associated superpotential~\eqref{quarticW}
and kink solution~\eqref{profile} one finds
\begin{equation}
 F(Z)=g v^3 \left. \left[ \tanh(\frac{g}{2}v\,(x-Z))-
      \frac{1}{3}\tanh^3(\frac{g}{2}v\,(x-Z))\right] \right|_0^L\; .
\end{equation}
This function has been plotted in Fig.~\ref{fig1} and it shows indeed
the general properties discussed above. The constant value of $F$
for $Z$ inside the interval is given by
\begin{equation}
 F(Z)\simeq 4g v^3/3\;\; {\rm for}\;\; Z\gg L_{\rm K}\;\;{\rm and}\;\;
 L-Z\gg L_{\rm K} \; .
 \label{F_inside}
\end{equation}
For later purposes, it is also useful to note the asymptotic
behaviour of $F$ for $Z\ll -L_{\rm K}$ which will be relevant after
a collision of the kink with the boundary at $x=0$. It is given by
\begin{equation}
 F(Z)\simeq 4 g v^3\,\left(1-e^{-2 gvL}\right)\, e^{2 gv\,Z}\;\; 
 {\rm for}\;\; Z\ll -L_{\rm K}\; .
\end{equation}
If the width of the kink $L_{\rm K}$ is much smaller than the
size $L$ of the interval, $L_{\rm K}\ll L$, this reduces to
\begin{equation}
 F(Z)\simeq 4g v^3\, e^{2 gv\,Z}\;\; {\rm for}\;\; Z\ll -L_{\rm K}
 \;\;{\rm and}\;\; L_{\rm K}\ll L\; .
 \label{F_assimpt}
\end{equation}

\vspace{0.4cm}

The equations of motion derived from the effective action~\eqref{Seff}
are simply
\begin{equation}
\frac{d}{dt}\left(F\dot{Z}^2\right)=0\; ,
\end{equation}
where the dot denotes the derivative with respect to time. This leads
to the first integral
\begin{equation}
 F\dot{Z}^2=u^2={\rm const}\; ,
\label{mod_en_cons}
\end{equation}
where $u$ is a constant. This relation, of course, represents energy
conservation. Together with the properties of $F$ stated above it
implies that $Z$ evolves with constant velocity in the interval $Z\in [0,L]$
and, once it has collided with the boundary at $x=0$,
accelerates and evolves towards $Z\rightarrow -\infty$ while $\dot{Z}$
diverges. Moreover, from the explicit form of $F$ one can show that
$Z$ reaches $-\infty$ in a finite time. Of course, the validity of the
effective theory~\eqref{Seff} when $Z$ and $\dot{Z}$ diverge after a
collision is at best doubtful. Although the effective theory provides
a good description of slow kink motion for finite values of $Z$, it
seems difficult, therefore, to draw any reliable conclusion about the
outcome of the collision based on this effective theory.  Three
possibilities are conceivable. The kink could be absorbed by the
boundary with its kinetic energy being completely converted into
radiation (that is, fluctuations of $\phi$ around one of its vacuum
states). In this case the charge on the absorbing boundary would have
changed by one unit between the initial and final state. It could be
inelastically reflected with part of its kinetic energy being transferred
to radiation or it could be elastically reflected. Then the boundary
charge remains unchanged. We will now analyse the full two-dimensional
theory~\eqref{Action} to determine which of these three possibilities
is actually realised.
\section{Perturbation spectrum}

It is instructive to examine how the eigenfunctions and eigenvalues of
small perturbations around the kink change with the modulus $Z$.  To
find the eigenvalues we must expand the field around the kink
configuration $\phi(x) = \phi_{\rm K}(x - Z)+ \varphi(x)$ and solve
the eigenvalue problem
\begin{equation}
\left[-\frac{d^2}{dx^2} + V''(\phi_{\rm K}) + 2W''(\phi_{\rm K})\delta(x)\right]\varphi = 
\omega^2\varphi 
\label{e:EigEqn}
\end{equation}
Here, we have neglected the contribution from the boundary at $x=L$
which is justified as long as the kink is sufficiently far away from
this boundary. This can always be achieved by making $L$ large.  We
would then expect that at large positive $Z$, the spectrum would be
asymptotically close to that of an isolated kink~\cite{Raj}.  For the
quartic potential~\eqref{quartic} one expects one zero mode with no
nodes, a bound state with one node and eigenvalue $\omega_1^2 =
3\mu^2/2$, and a continuum starting at $\omega_2^2 = 2\mu^2$, where
$\mu=gv/\sqrt{2}$. At large negative $Z$, the spectrum should be
asymptotically close to that of the vacuum, which also possesses a
nodeless zero mode and a continuum starting at $\omega^2 = 2\mu^2$, but
no bound state.

Although the free kink eigenfunctions and eigenvalues are known
exactly~\cite{Raj}, the non-trivial boundary condition, or
equivalently the extra $\delta$-function contribution to the
potential, change the problem.  As we have not been able to find a
solution to the eigenvalue problem in closed form, we resorted to a
simple numerical method.

Eq.~\eqref{e:EigEqn} was discretised on an $2N+1$-point interval $-L <
x <L$, using the lowest order discrete Laplacian
\begin{equation}
-\frac{d^2}{dx^2}(y) \to -\frac{\varphi_{i+1} - 2 \varphi_i +\varphi_{i-1}}{\Delta x^2},
\end{equation}
where $\Delta x = L/N$.  At the left (right) boundary the double
forward (backward) derivative was taken, which improves the accuracy
at the expense of generating two spurious eigenfunctions with large
negative eigenvalues.  The $\delta$-function was defined as $\delta(x)
\to \delta_{i0}/\Delta x$.  One should note that in defining the
problem for both positive and negative values of $x$, we will find
eigenfunctions which are both symmetric and antisymmetric under
reflections around $x=0$.  We are clearly interested only in the
symmetric ones.

\begin{figure}[htbp]
\begin{center}
\scalebox{0.5}{\includegraphics{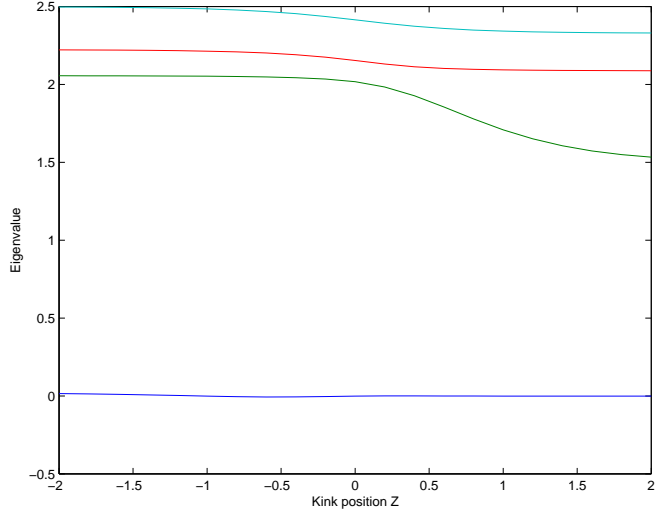}}
\caption{The lowest three eigenvalues $\omega^2$ as a function of kink position $Z$.  As $Z$ decreases, the bound state at $\omega^2 = 1.5$ moves towards the continuum at 
$\omega^2 = 2$.  In infinite volume it will remain bound, but
exponentially weakly.}
\label{f:EigZ}
\end{center}
\end{figure}

\begin{figure}[htbp]
\begin{center}
\scalebox{0.4}{\includegraphics{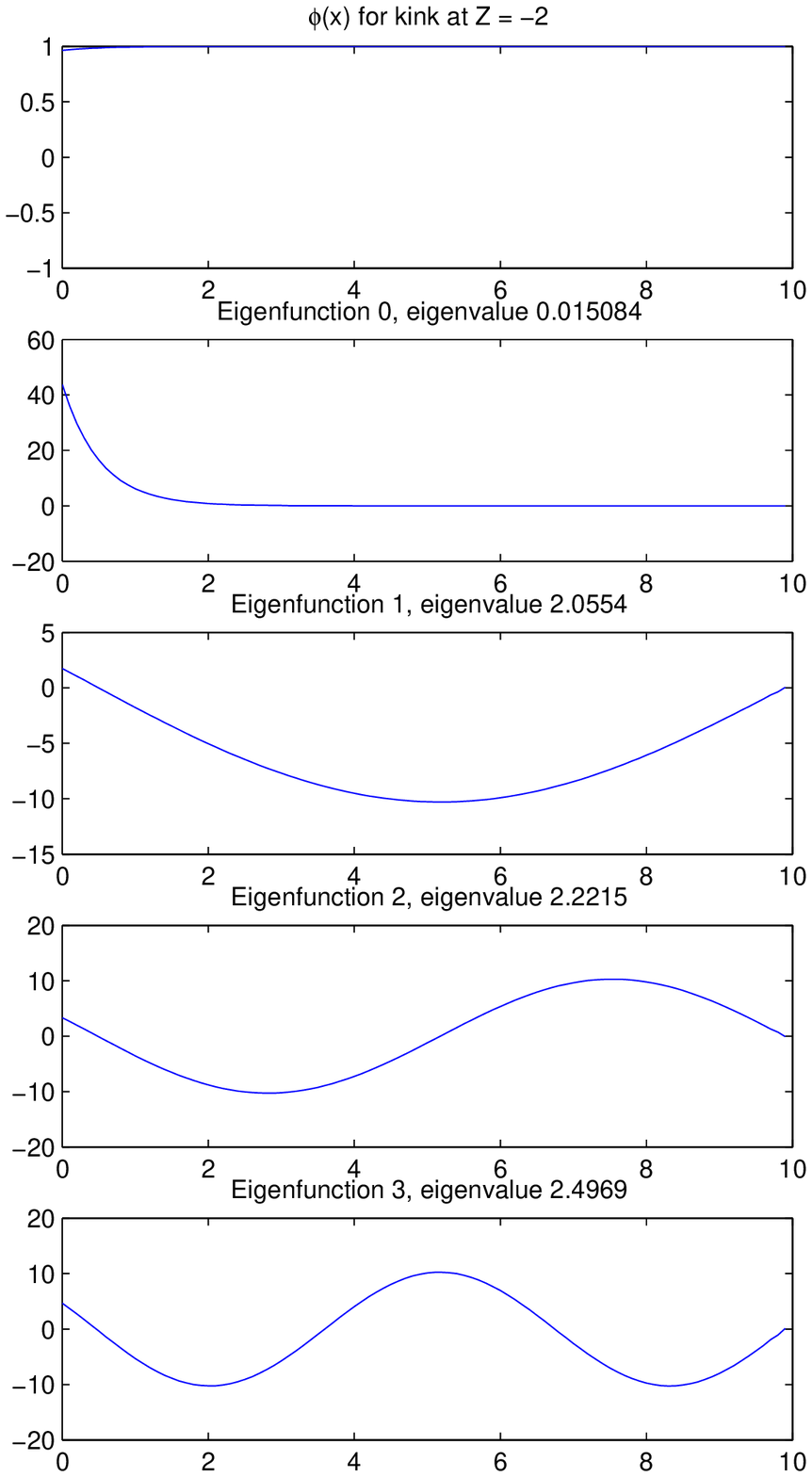}}
\scalebox{0.4}{\includegraphics{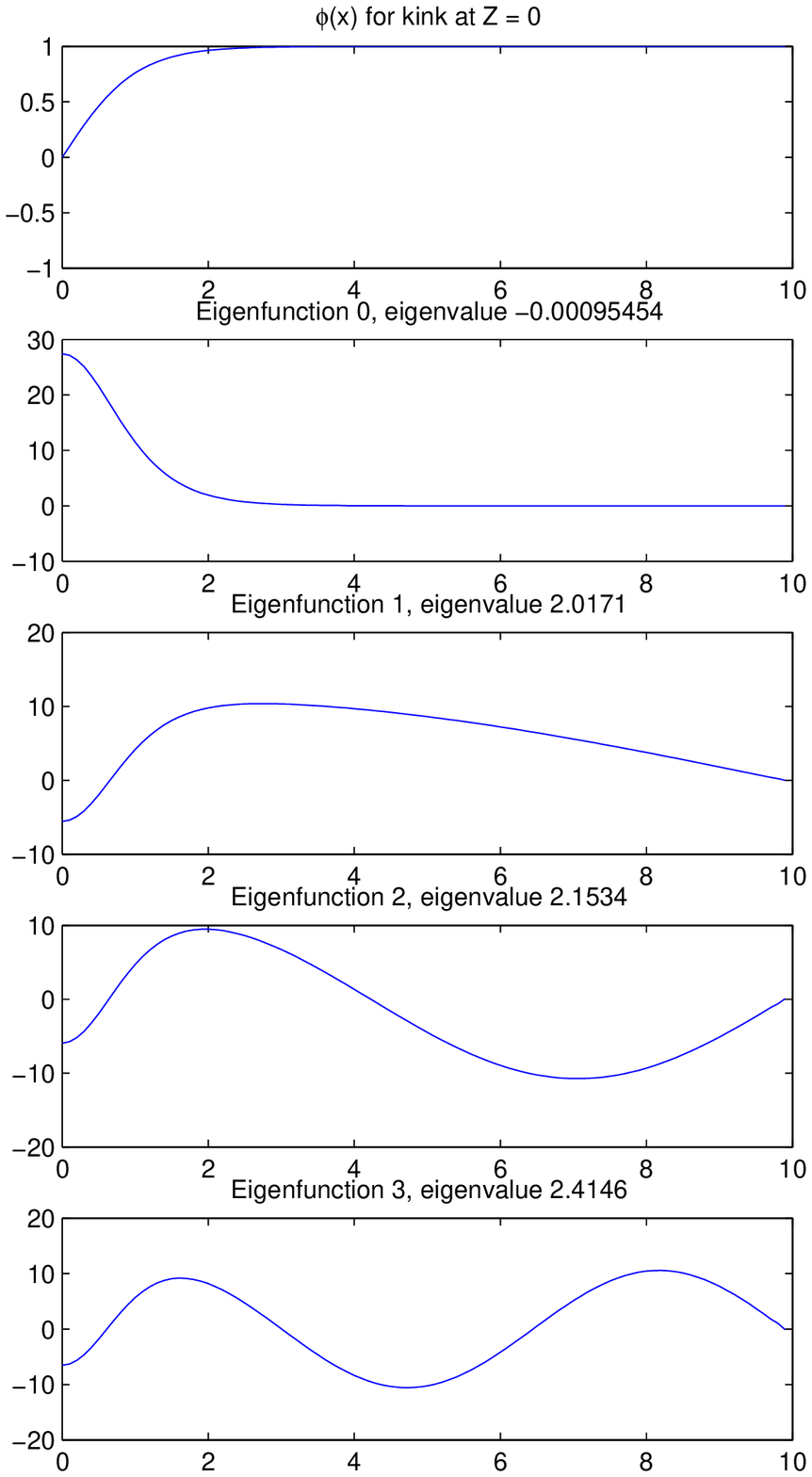}}
\scalebox{0.4}{\includegraphics{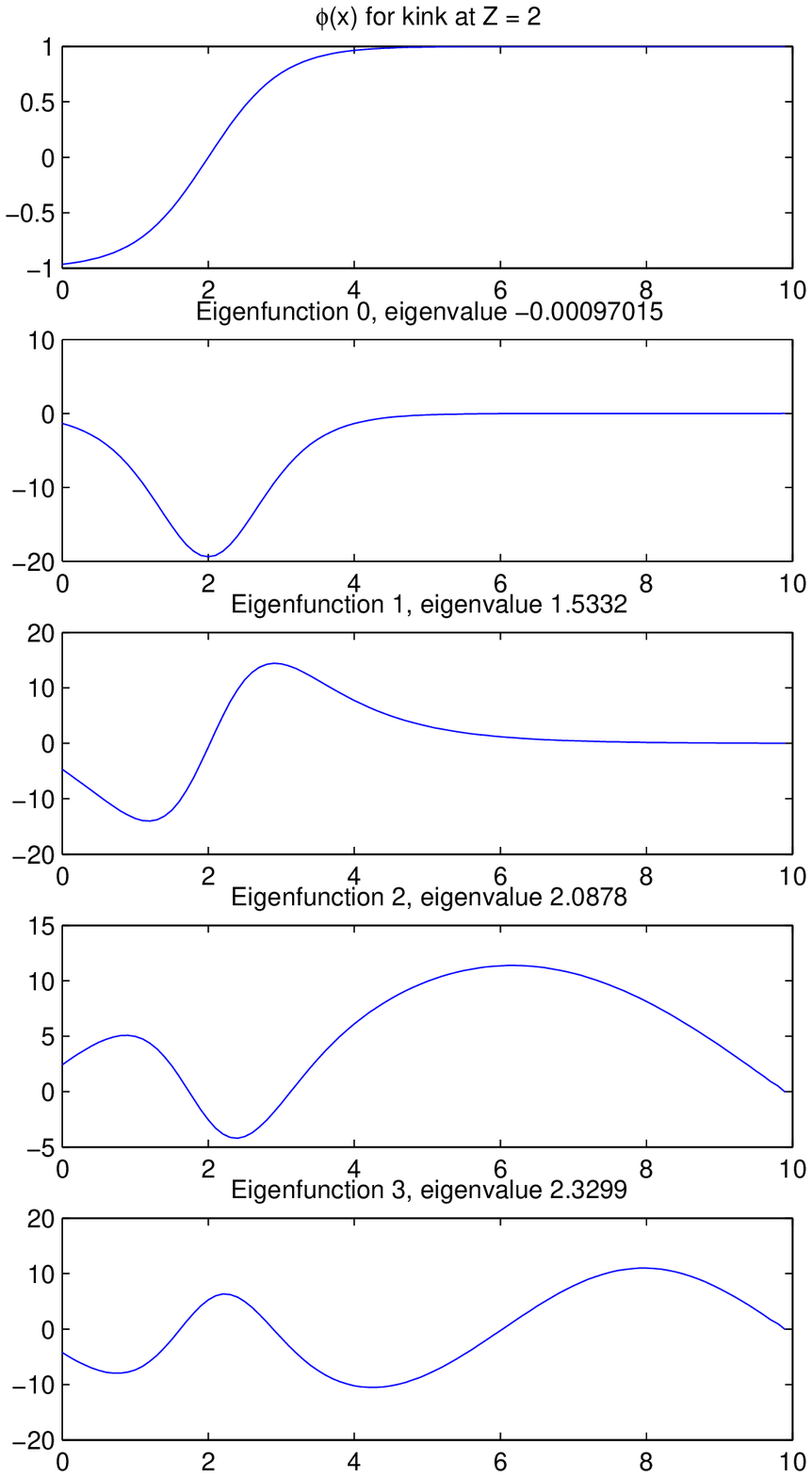}}
\caption{Graphs of the eigenfunctions corresponding to the lowest three eigenvalues for 
the kink at (a) $Z=-2$, (b) $Z=0$, and (c) $Z=2$.  At the top is a
     graph of the field $\phi$
     itself.}
\label{f:EigFun}
\end{center}
\end{figure}

The eigenfunctions and eigenvalues are easily found using {\sc
matlab}. We took $N=100$ and $L=10$, in units where $\mu=1$.  
The lowest three (0, 1 and 2
nodes) are shown for $-2 < Z <2$ in Fig.\ (\ref{f:EigZ}).  The
closeness of the lowest eigenvalue to 0 is a measure of the accuracy
of the procedure. The shapes of the eigenfunctions for $Z=-2$, $Z=0$
and $Z=2$ are shown in Fig.\ (\ref{f:EigFun}).

As we will see later, the existence of a zero mode for all values of
the kink position $Z$ is crucial for an understanding of the kink-boundary
collision. We will also find that it has important implications for the
creation of kinks due to spontaneous emission from the boundary.

\section{Numerical Simulation}
\label{numerics}
     
 As we have seen in the previous sections, a straightforward moduli
 calculation suggests that, as the kink collides with the boundary, it
 should cross it and accelerate towards infinity.  As a result, the
 field configuration inside the bulk will approach one of the true
 vaccua of the theory. We could then expect the kinetic energy of the
 kink to be converted into radiative excitations of the vacuum that
 would propagate away from the boundary. This reasoning suggests, the
 kink should be absorbed by the boundary as a result of the
 collision. In this section we will test this expectation by performing
 numerical simulations of the kink/boundary collision. As we will see,
 a more complex picture will emerge, showing how the moduli
 description fails to capture the essential features of the collision
 process.
     
 We model the bulk equation of motion Eq.~(\ref{EOM}) with the quartic
 potential~\eqref{quartic} on a $1D$ lattice using a second order
 discretization of the Laplacian operator, and propagate it in time
 according to a standard leapfrog algorithm. The implementation of the
 boundary condition is less straightforward, since it relates a
 function of the field to its spatial derivative. Taking as an example
 the boundary condition at $x=0$, we discretise it as
  \begin{equation}
  \frac{\phi_1-\phi_0}{\delta x}=W'(\phi_m),
 \label{BDdiscrete}
 \end{equation}
  where $\phi_m$ is the value of the field evaluated at some point
 in the interval $(0,\delta x)$. We allowed for different choices of
 the discretization scheme by writing $\phi_m=(1-G)\,\phi_0 +G\, \phi_1$,
 with $G$ a parameter lying between $0$ and $1$.
 Since the discretised spatial derivative is effectively evaluated in 
 the mid-point between lattice sites $i=0$ and $i=1$, the most
 natural option is to set $G=.5$ We confirmed that this choice is the
 most accurate one, by evolving in time a stationary kink configuration,
 with its core placed near the boundary. For $G>.5$ the kink is
 slightly repelled by the boundary, whereas for $G<.5$ the boundary
 has an attractive effect. The disadvantage of setting $G=.5$ is that
 it turns Eq.~(\ref{BDdiscrete}) into a non-linear implicit equation
 for $\phi_0$, which has to be solved numerically for each time-step. 
 This is clearly the case for any choice of $G$ other than $G=1$.
 On the other hand, for the lattice action obtained
 by a straightforward discretization of Eq.~(\ref{Action}), the 
 only energy conserving algorithm corresponds to $G=0$. For $G>0$
 we cannot rely on energy conservation as test of the adequacy
 of the choice of time-step. Of course all schemes conserve energy in
 the $\delta x\rightarrow 0$ limit.
 Since we are dealing with a one-dimensional system, we had no serious
 computational restrictions and we chose to work with $G=.5$. We have
 checked that, in the limit of small lattice spacing the three 
 values $G=0,.5$ and $1$ lead to the same results. The boundary 
 condition equation Eq.~(\ref{BDdiscrete}) was solved using an implementation
 of the Newton-Raphson method (see \cite{NumRec}).

 The initial conditions for the numerical evolution should model, as
 closely as possible, a kink in the bulk with a given velocity
 $\dot{Z}_i$ moving towards one of the boundaries. We set up the
 initial field configuration as $\phi_K(x-Z_i)$ using the kink profile
 in Eq.~(\ref{profile}), with $Z_i$ the initial core position far from
 both boundaries.  The initial field momentum is defined as
 $\Pi=-\dot{Z}_i\,\partial_x \phi_K$, where $\dot{Z}_i$ is the chosen
 initial kink velocity.  The simulations were run for a theory with
 $g=2\sqrt{2}$ and $v=1$, which implies a kink width of $L_{\rm
 K}=\frac{1}{2\sqrt{2}}$. We chose a box of physical length $L=16$,
 comfortably larger than the kink width. For high and mid-range kink
 velocities, the lattice parameters were set to $\delta x=.0125$ and
 $\delta t=.005$. Low collision velocities lead to a large total
 simulation time and a smaller time-step is needed to avoid error
 accumulation. For the lowest velocities we used $\delta x=.00125$ and
 $\delta t=.001$.
     
\begin{figure}
 \centering
 \includegraphics[height=15cm,width=15cm, angle=0]{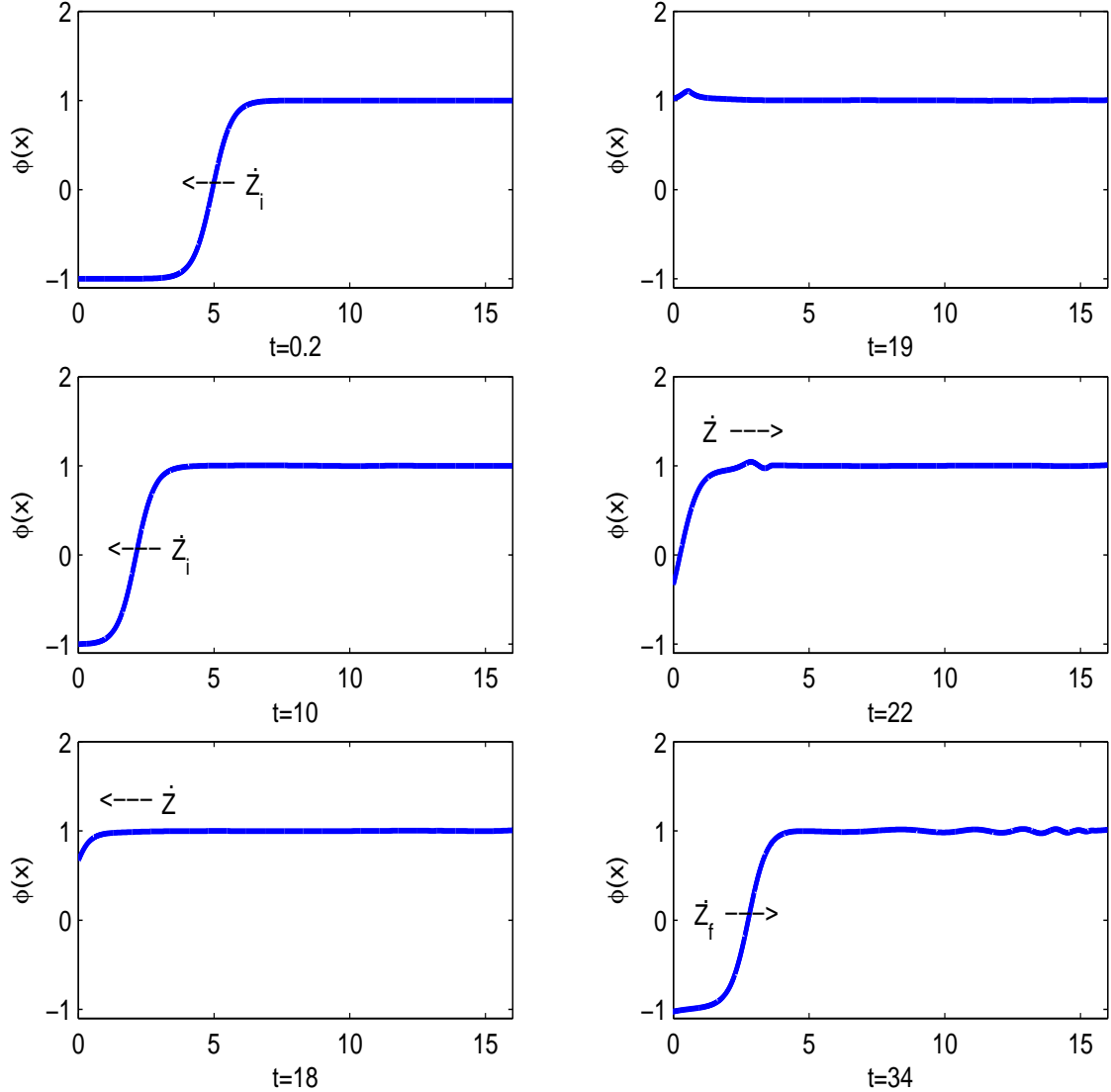}
 \caption{\emph{Kink collision for initial velocity $\dot{Z}_i=0.3$. In
  the first two snapshots, $t=.2$ and $t=10$, the kink moves towards
  the boundary with approximately constant velocity. For $t=18$ most of
  the core has already crossed the boundary, although the profile inside the 
  bulk remains relatively undistorted. For $t=19$ the field oscillates 
  around the true vacuum. These perturbations propagate away at the 
  speed of light, as seen in the next two plots. At $t=22$ the kink
  core has re-entered the bulk, and in the final plot we can see the final
  kink profile, moving away from the boundary with constant velocity.}}
 \label{fig2}     
 \end{figure}

  We explored a wide range of collision velocities, letting
 $\dot{Z}_i$ vary between $.01$ and $.75$ in steps of $.01$ In 
 Fig.~\ref{fig2} we can see a series of snapshots of $\phi$ for
 several evolution times in a typical run (in this case
 $\dot{Z}_i=.3$). During the first part of the evolution the
 kink moves towards the boundary with constant velocity. As its
 core starts moving into the boundary there is, as expected from
 the moduli picture, an increase in $\dot{Z}$. At some point the
 kink leaves the bulk completely and the field is left in the true 
 vacuum, oscillating rapidly in the vicinity of the boundary.
 The kink then ``returns'' into the simulation box, moving
 away from the boundary with decreasing velocity at first. 
 When its core is well inside the bulk the kink velocity oscillates
 around an average value $\dot{Z}_f$. This final value is
 always smaller than the initial velocity $\dot{Z}_i$, as some energy
 is lost into radiation during the collision. In the late time profiles in 
 Fig.~\ref{fig2} we can see these radiative perturbations moving 
 away from the boundary.
     
\begin{figure}
\centering
\includegraphics[height=9cm,width=9cm, angle=0]{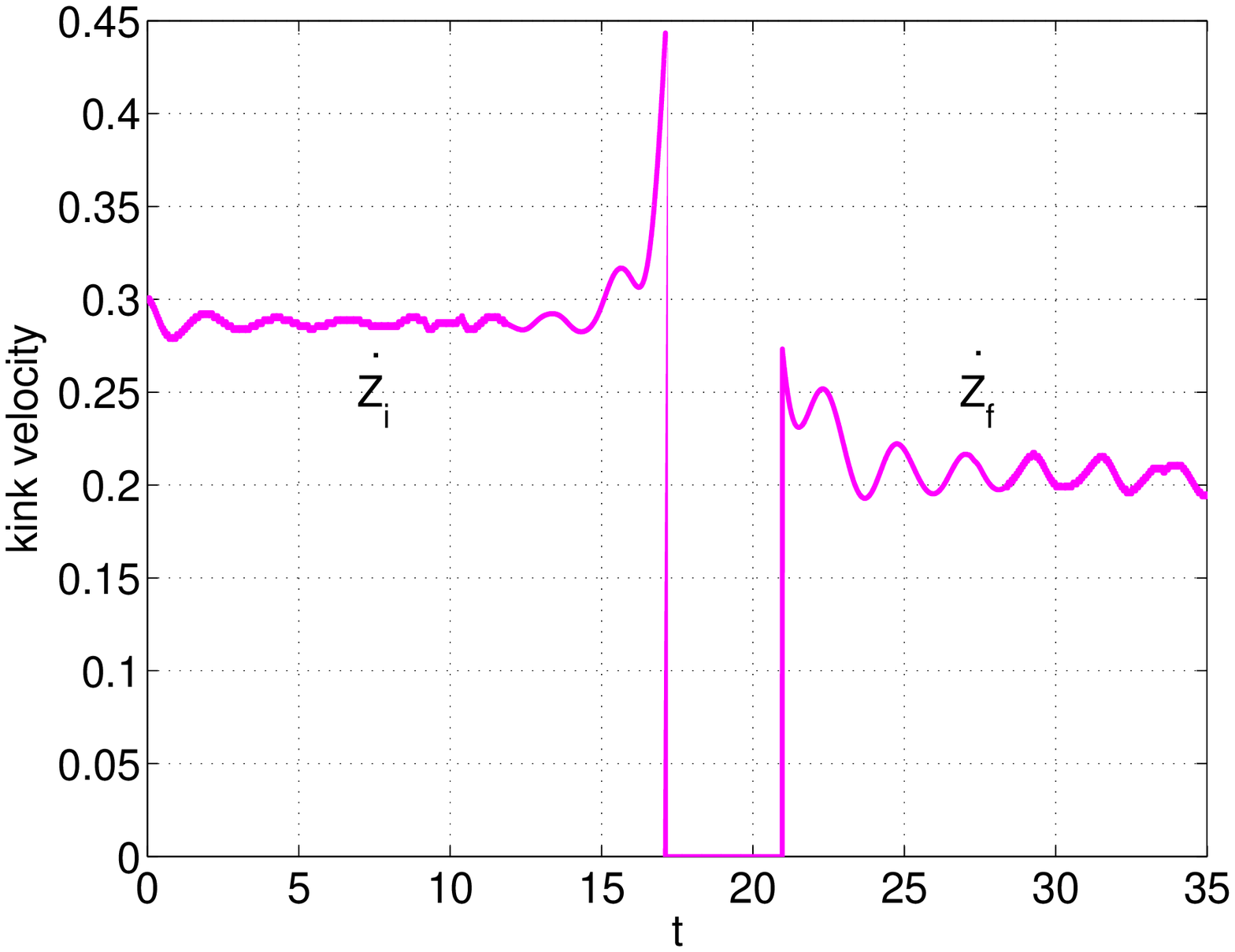}
\caption{\emph{Kink velocity versus time for initial velocity
$\dot{Z}_i=0.3$. As predicted by the moduli calculation, the kink 
accelerates as it approaches the boundary. After the collision
it returns into the bulk with a lower final velocity. During
the collision period, when the core is outside the bulk, the 
kink position cannot be determined by the zero of the field, and
we have defined the kink velocity as zero.}}
\label{fig3}
\end{figure}

 In Fig.~\ref{fig3} we have a plot of the kink velocity versus
 time. $\dot{Z}(t)$ was evaluated by numerical differentiation of
 the position of the core of the kink $Z(t)$. This in turn
 was defined as the position of the zero of the field, obtained
 by interpolating $\phi$ as its sign changes. The same features of the
 evolution can be seen, with $\dot{Z}(t)$ increasing as the kink
 approaches the boundary, and being reflected later on with a
 lower final velocity. Note that in both the beginning and end periods
 of the evolution, the velocity is not strictly constant, oscillating
 around an average value. This indicates that in both cases, 
 apart from the translational mode, other modes of the kink are
 excited. As far as the initial condition is concerned this was 
 expected since we are not using a Lorentz boosted kink, which would
 clearly not obey the boundary conditions. This effect is particularly
 visible for high kink velocities, both before and after the
 collision.
 
  Surprisingly, the overall qualitative pattern of the evolution did not
 change within the whole range of velocities observed. Even for very
 low velocities, where we expected the reasoning based on the moduli
 calculation to be applicable, the incoming kink was always reflected
 by the boundary after the collision. In the next few sections we will
 try to explain this behaviour and gain a deeper quantitative
 understanding  of the collision process.

\section{Collision as a vacuum perturbation}
\label{perturbation}
     
 As we have seen in Section~\ref{moduli}, the motion of the
 kink should be well described by the moduli equations
 of motion, up to the  point when the velocity of its core becomes
 too large. For low values of the initial velocity $\dot{Z}_i$,
 we expect this to happen when the kink core is outside the
 bulk and far from away from the boundary. That this is the
 case can be easily shown using the moduli energy conservation
 equation (\ref{mod_en_cons}).
 Assuming that the moduli approximation
 breaks down for velocities above a certain $\dot{Z}_{max}$. Then
 Eq.~(\ref{mod_en_cons}) implies that the corresponding core
 position $Z_{max}$ should satisfy
 \begin{equation}
   F(Z_{max}) = \frac{4}{3}g v^3 
  \left(\frac{\dot{Z}_i}{\dot{Z}_{max}}\right)^2\; .
 \end{equation}  
 Here we assumed the kink to be initially inside the bulk and
 far from its boundaries so that (\ref{F_inside}) applies.
 Clearly, the form of $F(Z)$ implies that for  $\dot{Z}_i$ small
 enough, $Z_{max}$ can be made arbitrarily large.
     
 If, when the maximal velocity is reached, we have $Z_{max}\ll-L_{\rm
 K}$, the field configuration in the bulk will be everywhere 
 very close to the vacuum $v$. This suggests that for low $\dot{Z}_i$,
 we could use the linearised equations of motion around $v$ to
 describe the evolution, after the moduli approximation breaks down.
 In this section we will use this approach to show that the kink
 bounces back inelastically after the collision, confirming the
 previous simulation results.

\subsection{Expansion basis for linearised equations of motion}

 As discussed above we perturb around the vacuum $v$ and write
 $\phi=\varphi+v$. For small $\varphi$ the field equation of
 motion reduces to
 \begin{equation}
  \ddot{\varphi}=\varphi'' - m^2 \varphi 
 \label{lin_eom}
 \end{equation}   
 where $m=g v$ in the case of the quartic potential. The non-trivial 
 contribution to the evolution will come from the boundary condition,
 which is given by
  \begin{equation}
  \varphi'(x_i,t) = (g/2)\,|\,\phi^2(x_i,t) - v^2|= m\,|\,\varphi(x_i,t)|
 \label{lin_bd} 
  \end{equation}
 In this section we will focus on a collision taking place at the left
 boundary, $x=0$, and effectively assume the bulk to be
 half-infinite. This should be a reasonable approximation as 
 long as the size of the bulk is larger than the collision time,
 so that effects from the radiation produced during collision reaching
 the opposite boundary can be neglected.
 Note that, as a consequence of the norm in (\ref{lin_bd}), the boundary 
 condition for $\varphi$ is non-linear. Nevertheless, for periods of
 the evolution during which $\varphi$ does not change sign at the
 origin, (\ref{lin_bd}) reduces to a linear equation. In particular
 we have
  \begin{eqnarray}
  \varphi'(0,t) = m\,\varphi(0,t),\,\,\,\,\,\,\,\,&{\rm if}&
     \varphi(0,t)>0,\label{bd_1}\\ 
 \varphi'(0,t) =- m\,\varphi(0,t),\,\,\,\,\,\,\,\,&{\rm
     if}&\varphi(0,t)<0 \label{bd_2}\; .
  \end{eqnarray}   
 This implies that we can still  
 expand arbitrary solutions of (\ref{lin_eom}) and (\ref{lin_bd})
 in terms of given mode functions, as long as we use a different
 basis for different periods of the evolution.
 We start with the case $\varphi(0,t)>0$. A complete basis
 of solutions of (\ref{lin_eom}) satisfying (\ref{bd_1}) is given
 by 
 \begin{equation}
  \varphi_k(x,t)=N_k\,e^{\pm i \omega
     t}\,\left[\cos(kx)+\frac{m}{k}\sin(kx)\right],\,\,\,\,\,\,w^2=k^2+m^2,\,\,\,\,k>0 \; .
  \label{basis_1}
 \end{equation}
 To obtain the normalisation coefficient we define the mode functions
 in a finite box of size $L$, leading to  $N_k^2 = 2 k^2/(L
 \omega^2)$. In the end of the calculation we will take
 $L\rightarrow \infty$. 
 As for $k$, it will have a discrete spectrum depending on the
 boundary conditions on the right end of the box. For simplicity
 we chose periodic boundary conditions, implying 
 $\cos(kL)+(m/k)\,\sin(kL)=1$. As the box size is taken to infinity
 this should not be relevant for the final result.
      
  In the case of negative field values $\varphi(0,t)$ the basis is given by
  \begin{equation}
  \varphi_k(x,t)=N_k\,e^{\pm i \omega t}\,\left[\cos(kx)-\frac{m}{k}\sin(kx)\right],\,\,\,\,\,\,w^2=k^2+m^2,\,\,\,\,k>0\label{basis_2}
 \end{equation} 
 with $\cos(kL)-(m/k)\,\sin(kL)=1$. Here $N_k^2$ is a normalisation which
 will be unimportant for the subsequent calculation.
 In addition, for the case of negative field values $\varphi(0,t)$ and
only for this case, there is also a zero-mass solution of the general
 form
  \begin{equation}
     \varphi_0(x,t)=(a+ b\,t)\,e^{-mx}
  \end{equation}    
 This zero mode corresponds to the tail of the kink when its core is
 far outside the bulk. It can be seen as the linearised version
 of the translational zero mode of the kink. As we will see, this is
 the essential feature that distinguishes the evolution in the two
 regimes, $\varphi>0$ and $\varphi<0$.
     
 \subsection{Linear solution}
     
 We now apply the above results to the kink/boundary collision
 situation. For simplicity, we define $t=0$ as the time when the
 moduli approximation breaks down. For this time, we
 must find the corresponding field configuration inside the
 bulk which can then be taken as initial condition for the equations
 of motion linearised around the vacuum.  
 When the kink core is very far from the boundary, the bulk field
 and its time derivative can be approximated by
 \begin{eqnarray}
    \phi(x,t)&=&v-2 v\, e^{m Z(t)}\, e^{-m x}\\
    \partial_t \phi(x,t)&=&-2 m v\, \dot{Z}(t)\, e^{m Z(t)}\, e^{-m x}
 \end{eqnarray}   
 We will assume that we can neglect the small perturbation around
 the vacuum and that effectively $\phi=v$ in the bulk at $t=0$. The
 derivative term on the other hand, remains finite even as $\dot{Z}(t)$
 diverges. Using energy conservation (\ref{mod_en_cons}) and
 (\ref{F_assimpt}) we obtain
 \begin{equation}
   4 m v^2\, e^{2 m Z}\, \dot{Z}^2 = F(Z_i)\,\dot{Z_i}^2
\end{equation}
which implies, using the asymptotic expansion for $F(Z)$, that
\begin{equation}
   e^{ m Z}\, \dot{Z} = \frac{\dot{Z}_i}{\sqrt{3}}\; . 
 \end{equation}
 This allows us to express
 the finite amplitude of the velocity profile at collision time in
 terms of the initial velocity of the kink. The perturbation field
 at $t=0$ is then defined as
  \begin{eqnarray}
  \varphi(x,0)&=&0\\
  \dot{\varphi}(x,0)&=&-\frac{2}{\sqrt{3}}\, m v\, \dot{Z}_i\, e^{-m
     x}
  \label{lin_in_con}
 \end{eqnarray} 
 These are now to be taken as initial conditions of the linearised
 equations of motion. Since $\dot{\varphi}>0$ we will be entering a
 phase of the evolution for which $\varphi(0,t)>0$ and we should 
 expand (\ref{lin_in_con}) in terms of the corresponding basis
  (\ref{basis_1}).
 The expansion coefficients are easily obtained (the result depends 
 only on integrals of products of exponentials and trigonometric
 functions) leading to 
 \begin{equation}
  \varphi(x,t)=\frac{4 A}{L}m
    \,\left(1-e^{-m L}\right)\,\sum_k
     \frac{k^2}{\omega^5}\,\sin(\omega
     t)\,\left[\cos(kx)+\frac{m}{k}\sin(kx)\right]
 \label{sol}
 \end{equation}
where $A=(2/\sqrt{3}) mv\,\dot{Z}_i$ is the amplitude of the exponential
tail in the initial velocity profile for the field. 
    
 Equation (\ref{sol}) is only valid as long as $\varphi(0,t)>0$. When
 the field becomes negative at the origin, the boundary condition
 changes sign and the solution should then be expanded in terms of the
 second basis (\ref{basis_2}). Let us define a ``reflection time''
 $t_R$, for which the solution becomes negative at the boundary, that
 is $\varphi(0,t_R)=0$. We can use (\ref{sol}) to calculate explicitly
 the mode amplitudes of $\varphi(x,t_R)=0$ in terms of the $\varphi<0$
 basis. The new ingredient now is the presence of the zero mode. All
 other modes are oscillatory, corresponding to the radiation that will
 propagate away from the boundary after the collision. The zero mode
 $\varphi_0$ on the other hand, will grow linearly in time, becoming
 the tail of the incoming kink. The situation will now mirror what
 happened before the collision, with the amplitude of the massless
 mode being related to the final velocity of the reflected kink
 $Z_f$. We define the zero mode amplitude as

 \begin{equation}
 a_0=\int dx\,\, \partial_t\varphi(x,t_R)\,N_0\,e^{-mx}
 \label{int}
 \end{equation}     
 where $N_0^2=2m/(1-e^{-2mL})$ is the corresponding normalisation
 factor. The time derivative of the field at $t_R$ will be given by:
 \begin{equation}
   \dot{\varphi}(x,t_R)= a_0\,N_0\,e^{-mx}+\,{\rm oscillatory\,\, terms}
 \end{equation}
 Since we are mainly interested in determining the final
 velocity of the reflected kink, we will not calculate explicitly
 the positive $k$ component of the field. We define the amplitude
 of the exponential term in $\dot{\varphi}(x,t_R)$ as $B=a_0\,N_0$,
 and in analogy with the 
 calculation for the kink before the collision, (\ref{lin_in_con})
 leads to
 \begin{equation}
   \dot{Z}_f=\frac{\sqrt{3}}{2mv}\,B\; . 
 \end{equation}
  A {\it reflection coefficient} can be defined as 
 $R=\dot{Z}_f/ \dot{Z}_i=B/A$, quantifying the elasticity of the
 collision. The amplitude $B$ can be easily obtained by performing the
 integral in Eq.~(\ref{int}), leading to the expression
  \begin{equation}
  R = \frac{B}{A}= N_0^2 \,\,\frac{2}{L}
    \,\left(1-e^{-m L}\right)^2\,4m^2\,\sum_k
     \frac{k^2}{\omega^{6}}\,\cos(\omega t_R)\,
 \label{B}
 \end{equation} 
 for the reflection coefficient where $\omega$ is given in
 Eq.~\eqref{basis_1}. Note that this solves the problem. We have
 obtained a formula for the reflection coefficient in terms of the
 reflection time $t_R$, which, in turn, is implicitly determined by
 Eq.~\eqref{sol} as the time satisfying $\varphi (0,t_R)=0$.
     
\subsection{Comparison with numerical data}
     
\begin{figure}
\centering
\includegraphics[height=9cm,width=9cm, angle=0]{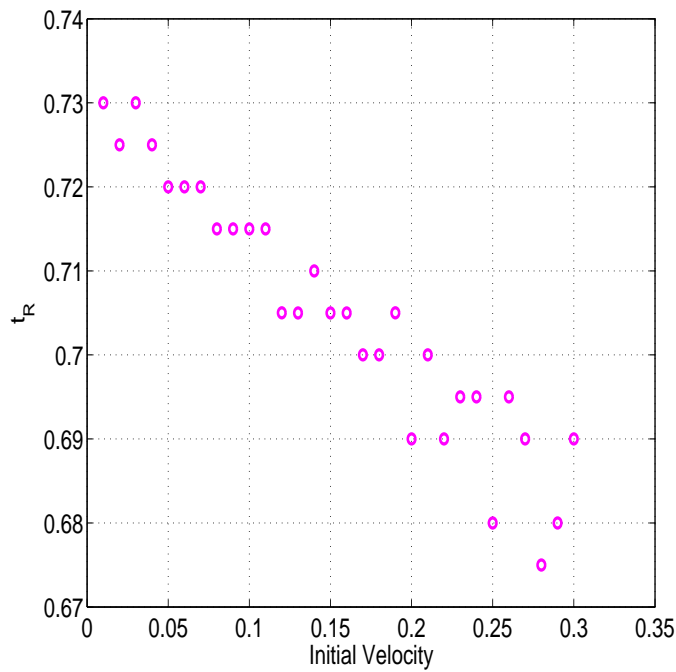}
\caption{\emph{Reflection time versus collision velocity
 for a model with $g=2\sqrt{2},\,\,v=1,\,\,m=2\sqrt{2}$. The 
 theoretical prediction (\ref{result}) for this choice of
 parameters is $t_R=0.73$. The reflection time $t_R$ was defined as
 the length of time for which $\phi(0,t)>v$ for each simulation.} }
\label{fig4}
\end{figure}

\begin{figure}
\centering
\includegraphics[height=9cm,width=9cm, angle=0]{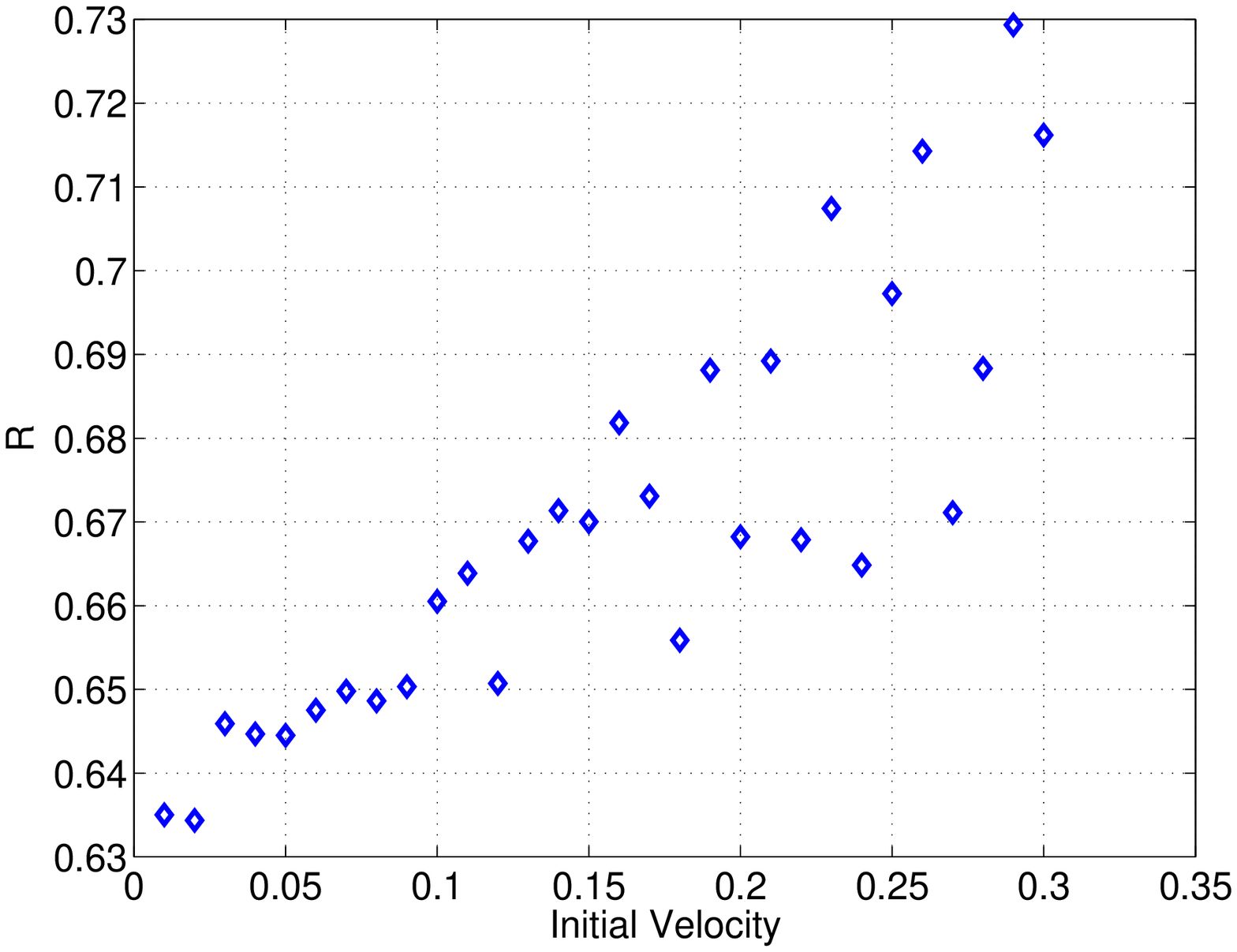}
\caption{\emph{Reflection coefficient versus collision velocity
 for $g=2\sqrt{2},\,\,v=1,\,\,m=2\sqrt{2}$. The (model independent)
 analytical prediction is $R=0.63$. Both the initial and final
 velocities were determined by fitting the position of the core
 of the kink (as defined in Section~\ref{numerics}) to a straight line
 in the relevant regimes.}}
\label{fig5}
\end{figure}

 To evaluate exact numerical values it is convenient to approximate
 the series obtained in the previous section by integrals. There is a slight 
 subtlety involved, related to the fact that we must add over all
 values of $k$ that solve the periodic boundary condition
 $\cos(kL)+(m/k)\,\sin(kL)=1$. This equation is solved by $k=2n\pi/L$,
 and by an extra non-trivial solution in every interval 
 $(2n\pi/L,2(n+1)\pi/L)$. This second solution approaches $k=2n\pi/L$
 as $n\rightarrow \infty$. 
 In replacing sums  over $k$  by integrals we must take into consideration
 the fact that there are two $k$'s in each interval of length $2\pi/L$ 
 and use $(\pi/L) \sum_k \rightarrow \int dk$ (rather than
 $(2\pi/L) \sum_k \rightarrow \int dk$). Using the continuum version
 of (\ref{B}) we obtain for the reflection coefficient     
 \begin{equation}
   R=\frac{16 m^3}{\pi} \,\int_0^\infty dk\,\frac{k^2}{\omega^6}\,\cos(\omega t_R)
 \end{equation}
 The reflection time $t_R$ is given implicitly by
 \begin{equation}
  \int_0^\infty dk\,\frac{k^2}{\omega^5}\,\sin(\omega t_R)=0
 \end{equation}
  Both integrals can easily be re-scaled and evaluated numerically,
 leading to the final results:
   \begin{equation}
      t_R=\frac{2.06}{m},\,\,\,\,\,\,\,R=0.63
   \label{result}
   \end{equation}
 These compare very well with the simulation results in the limit of
 low collision velocities, as illustrated in Figs.~\ref{fig4} and
 \ref{fig5}. As the initial kink velocity increases deviations from
 the analytical result can be observed. This is to be expected,
 since for higher velocities the moduli approximation should 
 break down earlier, making equation (\ref{lin_in_con}) a worse
 guess for the form of the initial conditions for the linear regime.
 Also, we should expect corrections to the linear approximation
 as the field starts probing the non-linearities of
 the potential during the collision, for high $\dot{Z}_i$. Still,
 in the particular system simulated, the numerical results remain
 within 10\% of the analytical prediction for velocities up to
 $\dot{Z}_i=0.3$.

  It is interesting to note that the equation (\ref{result}) depends
 only on the mass of the perturbed theory around the vacuum. In fact,
 it is easy to see that the derivation leading to this result could
 be reproduced for a general potential with non-vanishing second
 order derivative at the true vacuum. In particular, for this class
 of potentials, the reflection coefficient should be model
 independent. We confirmed this remarkable result by simulating
 a few collision velocities for a sine-Gordon model. As expected,
 we observed $R=0.63$.

\section{Moduli space interpretation of the kink-boundary collision}

Can we find an interpretation for our results in terms of a suitable
modification of the one-dimensional moduli-space approximation described in
Section~\ref{moduli}? As before, we focus on a collision with the
boundary at $x=0$. In view of the effective action~\eqref{Seff}, it
seems natural to introduce a new field $\zeta$ with a canonically
normalised kinetic term by setting
\begin{equation}
 \frac{d\zeta}{dZ}=\pm\sqrt{F}\; ,\qquad
 \zeta (Z) = \pm\int_{-\infty}^Zd\tilde{Z}\sqrt{F(\tilde{Z})}\; .
 \label{zeta}
\end{equation}
Note that the integral in this definition is indeed finite which
corresponds to the fact that $Z$ reaches $-\infty$ in a finite
time. For a positive (negative) sign in the above definition
the full range of $Z$ values is mapped into positive (negative)
$\zeta$ values while the point $Z=-\infty$ is mapped to $\zeta =0$
for both signs. The effective action~\eqref{Seff} now takes the canonical
form
\begin{equation}
 S_{\rm eff} = \int dt\, \dot{\zeta}^2 \label{z1}
\end{equation}
and leads to the first integral
\begin{equation}
 \dot{\zeta} = V = {\rm const}\; .
\end{equation}
Let us discuss the collision process in terms of the new field
$\zeta$. We start off at some finite, positive value $\zeta$,
corresponding to the kink being located between the boundaries, and
move with a constant, negative velocity towards $\zeta = 0$.
When $\zeta$ reaches some smaller, positive value, corresponding
to the position of the boundary at $x=0$, the collision takes
place. From there $\zeta$ continues to evolve to zero and,
finally, into negative values. Nothing special happens at
$\zeta =0$ which corresponds to the formerly problematic 
point $Z=-\infty$. We would now like to re-interpret this
evolution of $\zeta$ in terms of our original field $Z$. 
Note that, from Eq.~\eqref{zeta} we have
\begin{equation}
 \dot{Z} = \left\{\begin{array}{rrr}+\frac{V}{\sqrt{F}}&{\rm for}&
           \zeta >0\\
           -\frac{V}{\sqrt{F}}&{\rm for}&\zeta <0\end{array}\right.\; .
\end{equation}
This means crossing of $\zeta =0$ after a collision implies a
sign change in the velocity of $Z$ and, hence, a reflection
of the kink. This is precisely what we have found in the
full two-dimensional theory. An alternative way of stating this
is to note that $\zeta$ is really defined on $R/Z_2$. Here
the $Z_2$ identifies $\zeta$ and $-\zeta$ which correspond to
the same field value $Z$. In fact, upon including the other
boundary the $\zeta$ moduli space is, perhaps not surprisingly,
given by $S^1/Z_2$. Note, however, the fix points of this moduli
space orbifold do not quite correspond to the fix points of the
space-time orbifold (which are at $x=0,L$) but rather to the
points ``at infinity", that is, $Z=\pm\infty$.


There are two obvious conclusions about the collision process
which we can deduce from our simple picture.
First, the absolute values of the kink velocity, $|V|$,
before and after the collision are the same, that is, the
reflection is elastic and the reflection coefficient is $R=1$.
This is in obvious disagreement with our two-dimensional results
which showed the reflection is inelastic even at low velocities with
a reflection coefficient $R\simeq 0.63$. This disagreement arises
because the modified moduli space picture still does not correctly describe
the system for a short time period, corresponding to the reflection
time $t_R$, when the modulus $\zeta$ 
is not defined.
During this
short period, the additional radiation is created, a process which
clearly cannot be described by our effective moduli space theory
with a single degree of freedom.  

Secondly, the total time $t_{\rm refl}$ between the two crossings
of the core of the kink with the boundary is given by
\begin{equation}
 t_{\rm refl} = \frac{2}{V}\int_{-\infty}^0\sqrt{F(Z)}dZ\; .
 \label{trefl}
\end{equation}
Note, this time is different from the reflection time $t_R$
defined earlier and, for low velocities, we have $t_{\rm refl}\gg t_R$.
While $t_{\rm refl}$ is the total time the kink vanishes ``behind" the boundary,
$t_R$ is the time the moduli space approximation breaks down when the
kink is far outside at $Z\rightarrow -\infty$ or $\zeta\simeq 0$.
Comparison with the numerical simulations shows that the above formula
approximates $t_{\rm refl}$ well for low velocities with the typical
discrepancy being of order $t_R$. This is understandable since the
main contribution to the integral in Eq.~\eqref{trefl} comes from the region
$Z\simeq 0$ around the boundary where the moduli space approximation is
still valid.

Another conclusion from Eq.~\eqref{trefl} is that the kink can disappear
behind the boundary for an arbitrarily long time for sufficiently low
initial velocity. However, as we have seen, it eventually always
reemerges and is reflected back into the bulk. 


\section{Conclusions}
\label{conclusions}

 Combining the numerical and analytical results discussed above,
 a consistent picture of the kink/boundary collision process
 begins to emerge. We found that for most of the evolution the system is well
 described by an effective theory for the kink position modulus. As the kink 
 leaves the bulk and the modulus diverges, there is a short period of time 
 (the reflection time $t_R$, in the notation of Section~\ref{perturbation})
 for which this approximation breaks down. During this stage, the field
 in the bulk takes values outside the interval $(-v,v)$ between the two
 minima of the potential and, as a consequence, the moduli approximation
is not defined.
 Solving the full field theory for this period of the evolution, we showed that 
 the kink is always inelastically reflected back into to the bulk. After
 the collision the kink moves in the bulk following once again the moduli
equations of motion. We presented an analytical approach which describes
the kink reflection as a fluctuation of the scalar field around its vacuum
state. This picture led to analytic expressions for the reflection coefficient
$R$ (the ratio of initial and final kink velocity) and the reflection time $t_R$.
These analytical results turned out to be in good agreement with the
numerical simulations for low kink velocities, thereby
supporting our analytical approach. Remarkably, the reflection coefficient
and, hence, the fraction of energy lost into radiation during a collision,
turned out to be model-independently given by $R\simeq 0.63$.

Although we have shown that the reflection of the kink can be modelled by
a formal continuation of the moduli space, some aspects of the collision,
such as its inelasticity, and an extra time delay $t_R$, 
are not correctly reproduced within this moduli
space approach. So far we have not been able to find an alternative
description of the collision that does not require taking into account
the full set of degrees of freedom of the two-dimensional field theory.
It is possible that a more adequate moduli space description can be
achieved by a different continuation of the moduli space but clearly
more work is needed to settle this question.

These results may be disappointing at first as they seem to imply 
that the number of kinks in the bulk and the charges on the boundaries
are always conserved during collisions. Initially we had expected to
find examples of collisions for which the kink would be absorbed by
the boundary, leading to a change of the boundary charge~\cite{Antunes:2002hn}.
In the related M-theory model~\cite{Lukas:1998yy,Antunes:2002hn} such
a change in the boundary charge indicates a change in the topology of
the M-theory background and is, therefore, of particular interest.

However, on closer examination, the mechanism behind the reflection
of the kink reveals a much richer potential for processes changing the
boundary charges. As shown in Section~\ref{perturbation}, after the
collision the kink grows back into the bulk as the exponential zero mode around the vacuum becomes excited.
If we now consider a general arbitrary perturbation of the vacuum, it is very
likely that it will have a non-vanishing component in the direction of the zero
mode. This will lead to a kink forming and moving into the core, changing
by $-1$ the charge of the boundary. The somewhat unexpected conclusion
seems to be that it is reasonably easy to extract a kink from the
boundaries. In fact it is easy to show that this process is not inhibited
by a lower energy threshold. When a kink forms, the change in the bulk
energy corresponding to its core is compensated for by an equal change
in the boundary energy terms. The energy of the initial perturbation is
then converted exclusively into kinetic energy of the incoming kink.
This implies that kinks with arbitrarily low velocities can always be
obtained from field configurations arbitrarily near the vacuum.
In this sense emission of kinks from the boundaries which change the
boundary charges becomes unavoidable, in practise.
     
 \begin{figure}
 \centering
 \includegraphics[height=15cm,width=15cm, angle=0]{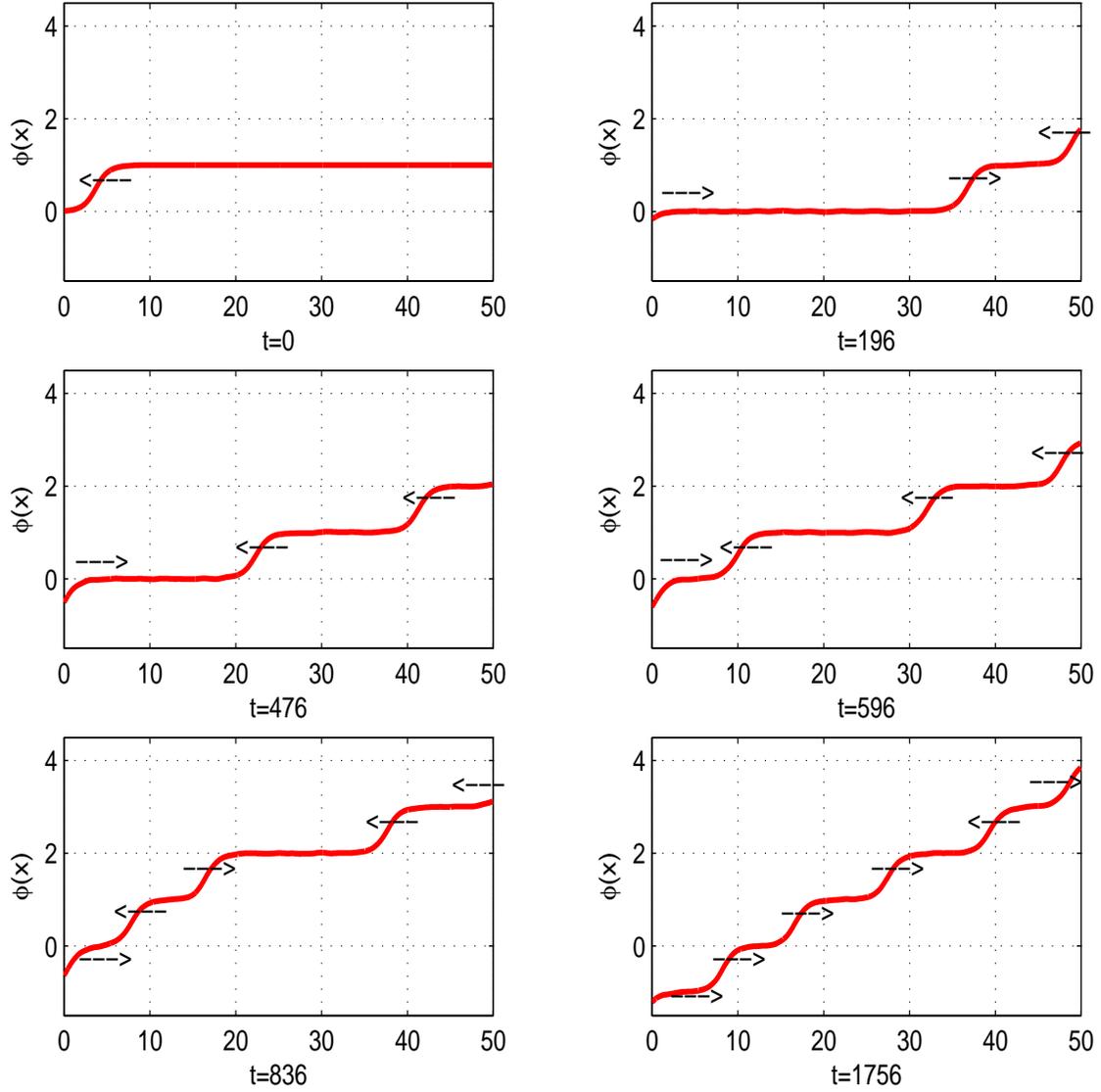}
 \caption{\emph{Initial single colliding kink gives rise to
  configurations with increasingly higher kink number. A sine-Gordon
  potential~\eqref{sg} with $m=1$ and $\lambda =1$ was used. The y-axis
 is normalised such that the vacuum points correspond to integers. The initial
 kinetic energy of the single kink is distributed amongst the final
 kinks that have lower velocity.}}
 \label{fig6}     
 \end{figure}
       
 Let us illustrate these ideas by looking at the very long time limit
 of the kink/boundary collision process. Here we will follow
 \cite{Antunes:2002hn} and consider a theory with a periodic potential
 $V(\phi)$. In the situations considered so far the field values were
 always near the two vaccua $-v$ and $v$, and the large $\phi$
 structure of the potential was never relevant. This is not the case,
 as we will see, for considerably long times after the collision.  As
 we have discussed, when the kink collides with the left boundary some
 radiation is produced which propagates away in the direction of the
 opposite right boundary (and vice versa). As the vacuum in the right
 boundary becomes perturbed, we can expect as before, that the
 corresponding zero mode becomes excited. From our arguments above
 this zero mode should with time grow into an additional, left-moving kink.
 Something similar could be expected to happen in the left boundary,
 after the reflected kink moves away into the bulk. Since we expect
 some small perturbation around the vacuum there, another (probably
 very slow) right-moving kink should start forming. For a periodic
 potential and very long times this process would repeat itself
 several times, with increasing numbers of kinks being extracted from
 both boundaries. That this seems to be indeed the case has been confirmed in
 Fig.~\ref{fig6}. For a sine-Gordon model with potential
\begin{equation}
 V(\phi)=-\frac{m^4}{\lambda}\left[\cos\left(\frac{\sqrt{\lambda}}{m}\phi
 \right)-1\right] \label{sg}
\end{equation}
and parameters $m=1$, $\lambda=1$ we observed the
 formation of five kinks from an initial single-kink collision. As
 expected, the kinetic energy of the initial kink is distributed
 amongst the final kinks that move with much lower velocities.
   
  As an extra test we evolved a number of near vacuum configurations
 with zero bulk charge and observed the formation of a number of kinks
 at both boundaries, as expected. The natural question at this point
 is which mechanism would cause the kink formation process to stop
 eventually.  One possibility is that, as their velocities become
 increasingly small, a kink gas would form, its density being limited
 by the repulsive kink-kink interaction. We can conjecture that for a
 very large family of initial conditions a gas of stacked branes
 should form, with density closely related to the energy of the
 initial field configuration. If this happens to be the case, the
 final charge configuration, in particular the final boundary charges
 of the system, would be selected by its total energy. Clearly a
 deeper understanding, both in numerical and analytical terms, of the
 system is needed in order to clarify these intriguing possibilities.


\vspace{1cm}

\noindent
{\Large\bf Acknowledgements}\\
A.~L.~is supported by a PPARC Advanced Fellowship.
N.~D.~A.~is supported by a PPARC Post-Doctoral Fellowship.



\begin{thebibliography}{99}

\bibitem{Antunes:2002hn}
N.~D.~Antunes, E.~J.~Copeland, M.~Hindmarsh and A.~Lukas,
``Kinky brane worlds,''
arXiv:hep-th/0208219.

\bibitem{Lukas:1998yy}
A.~Lukas, B.~A.~Ovrut, K.~S.~Stelle and D.~Waldram,
``The universe as a domain wall,''
Phys.\ Rev.\ D {\bf 59} (1999) 086001
[hep-th/9803235].

\bibitem{Ellis:1998dh}
J.~R.~Ellis, Z.~Lalak, S.~Pokorski and W.~Pokorski,
``Five-dimensional aspects of M-theory dynamics and supersymmetry  breaking,''
Nucl.\ Phys.\ B {\bf 540} (1999) 149
[hep-ph/9805377].

\bibitem{Lukas:1998tt}
A.~Lukas, B.~A.~Ovrut, K.~S.~Stelle and D.~Waldram,
``Heterotic M-theory in five dimensions,''
Nucl.\ Phys.\ B {\bf 552} (1999) 246
[hep-th/9806051].

\bibitem{Lukas:1998hk}
A.~Lukas, B.~A.~Ovrut and D.~Waldram,
``Non-standard embedding and five-branes in heterotic M-theory,''
Phys.\ Rev.\ D {\bf 59} (1999) 106005
[hep-th/9808101].

\bibitem{Brandle:2001ts}
M.~Brandle and A.~Lukas,
``Five-branes in heterotic brane-world theories,''
Phys.\ Rev.\ D {\bf 65} (2002) 064024
[arXiv:hep-th/0109173].

\bibitem{Witten:1996gx}
E.~Witten,
``Small Instantons in String Theory,''
Nucl.\ Phys.\ B {\bf 460} (1996) 541
[hep-th/9511030].

\bibitem{Ganor:1996mu}
O.~J.~Ganor and A.~Hanany,
``Small $E_8$ Instantons and Tensionless Non-critical Strings,''
Nucl.\ Phys.\ B {\bf 474} (1996) 122
[hep-th/9602120].


\bibitem{IntegBound}
S.~Ghoshal and A.~B.~Zamolodchikov,
Int.\ J.\ Mod.\ Phys.\ A {\bf 9}, 3841 (1994)
[Erratum-ibid.\ A {\bf 9}, 4353 (1994)]
[arXiv:hep-th/9306002];
S.~Ghoshal,
Int.\ J.\ Mod.\ Phys.\ A {\bf 9}, 4801 (1994)
[arXiv:hep-th/9310188];
H.~Saleur, S.~Skorik and N.~P.~Warner,
Nucl.\ Phys.\ B {\bf 441}, 421 (1995)
[arXiv:hep-th/9408004].

\bibitem{DeWolfe:1999cp}
O.~DeWolfe, D.~Z.~Freedman, S.~S.~Gubser and A.~Karch,
``Modeling the fifth dimension with scalars and gravity,''
Phys.\ Rev.\ D {\bf 62} (2000) 046008
[hep-th/9909134].

\bibitem{Copeland:2001zp}
E.~J.~Copeland, J.~Gray and A.~Lukas,
``Moving five-branes in low-energy heterotic M-theory,''
Phys.\ Rev.\ D {\bf 64} (2001) 126003
[hep-th/0106285].

\bibitem{Raj} R.~Rajaraman, ``Solitons and Instantons,''
(North-Holland, Amsterdam, 1982).
     
\bibitem{NumRec} W. H. Press, B. P. Flannery,
     S. A. Teukolsky and W. T. Vetterling, {\it Numerical Recipes
     in C: The Art of Scientific Computing}, (Cambridge University
     Press, Cambridge, 1990).



\end{thebibliography}
\end{document}